\documentclass[english,showpacs,preprint,superscriptaddress]{revtex4-1}
\usepackage{lmodern}
\usepackage{lmodern}
\usepackage[T1]{fontenc}
\usepackage[latin9]{inputenc}
\usepackage{amsmath}
\usepackage{amssymb}
\usepackage{esint}

\makeatletter

\@ifundefined{textcolor}{}
{%
 \definecolor{BLACK}{gray}{0}
 \definecolor{WHITE}{gray}{1}
 \definecolor{RED}{rgb}{1,0,0}
 \definecolor{GREEN}{rgb}{0,1,0}
 \definecolor{BLUE}{rgb}{0,0,1}
 \definecolor{CYAN}{cmyk}{1,0,0,0}
 \definecolor{MAGENTA}{cmyk}{0,1,0,0}
 \definecolor{YELLOW}{cmyk}{0,0,1,0}
}

\usepackage{amsthm}\usepackage{latexsym}\usepackage{bm}\usepackage{amsfonts}

\usepackage[english]{babel}

\makeatother

\begin{document}

\title{Field theory and weak Euler-Lagrange equation for classical particle-field
systems}

\author{Hong Qin}

\affiliation{Department of Modern Physics, University of Science and Technology
of China, Hefei, Anhui 230026, China}

\affiliation{Plasma Physics Laboratory, Princeton University, P.O. Box 451, Princeton,
New Jersey 08543}

\author{Joshua W. Burby}

\affiliation{Plasma Physics Laboratory, Princeton University, P.O. Box 451, Princeton,
New Jersey 08543}

\author{Ronald C. Davidson}

\affiliation{Plasma Physics Laboratory, Princeton University, P.O. Box 451, Princeton,
New Jersey 08543}
\begin{abstract}
It is commonly believed as a fundamental principle that energy-momentum
conservation of a physical system is the result of space-time symmetry.
However, for classical particle-field systems, e.g., charged particles
interacting through self-consistent electromagnetic or electrostatic
fields, such a connection has only been cautiously suggested. It has
not been formally established. The difficulty is due to the fact that
the dynamics of particles and the electromagnetic fields reside on
different manifolds. We show how to overcome this difficulty and establish
the connection by generalizing the Euler-Lagrange equation, the central
component of a field theory, to a so-called \textit{weak }form. The
weak Euler-Lagrange equation induces a new type of flux, called the
weak Euler-Lagrange current, which enters conservation laws. Using
field theory together with the weak Euler-Lagrange equation developed
here, energy-momentum conservation laws that are difficult to find
otherwise can be systematically derived from the underlying space-time
symmetry. 
\end{abstract}


\maketitle

\section{Introduction }

It has been widely accepted as a fundamental principle of physics
that energy-momentum conservation of a classical or quantum system
is due to the underlying space-time symmetry that the system admits.
However, for classical systems with particles and self-generated interacting
fields, the connection between energy-momentum conservation and space-time
symmetry has only been cautiously suggested \cite{Landau75} and has
not been formally established. Examples of such classical particle-field
systems include charged particles in an accelerator or a magnetic
confinement device interacting through the self-consistent electromagnetic
fields \cite{Davidson01-23}. To understand and overcome this difficulty,
we need to examine the details of the field theory. In the standard
field theory, one writes down a Lagrangian density $L$, and the associated
Euler-Lagrange equation determines the dynamics of the system. When
the Euler-Lagrange equation is satisfied, a symmetry condition is
equivalent to a conservation law. This is of course the celebrated
Noether's theorem \cite{Noether1918,Olver93-242}. It is surprising
to find out for classical particle-field systems that the standard
Euler-Lagrange equation does not hold anymore. This is because the
dynamics of the particles and the electromagnetic fields reside on
different manifolds. The electromagnetic fields are defined on the
space-time domain, whereas the particle trajectories as a field are
only defined on the time-axis. This is why the link between the symmetry
and conservation law breaks down for these systems. This unique feature
has not been discussed before, and it makes a significant difference
in the formulation of the field theory presented here. What we have
discovered is that when the standard Euler-Lagrange equation breaks
down, the field equations of these systems assume a more general form
that can be viewed as a weak Euler-Lagrange equation. It a pleasant
surprise to find out that this weak Euler-Lagrange equation can also
link symmetries with conservation laws as in the standard field theory,
where the regular Euler-Lagrange equation provides the link. The difference
is that the weak Euler-Lagrange equation induces a new type of current
(unknown previously), called the weak Euler-Lagrange current, in conservation
laws, in addition to the Noether current for the standard field theory.
For many classical particle-field systems, such as particles interacting
through electrostatic potentials \cite{Tonks29,Davidson01-23} or
attracting Newtonian potentials \cite{Rein97,Andreasson05}, energy-momentum
conservation laws are difficult to find. Using the field theory with
the weak Euler-Lagrange equation developed here, energy-momentum conservation
laws can be systematically derived from the underlying space-time
symmetries. 

This paper is organized as follows. In Sec.\,\ref{sec:CPFS}, the
classical particle-field systems and the difficulty of establishing
the connections between symmetries and conservation laws are introduced.
The weak Euler-Lagrange equation and its role in estabilishing conservation
laws are given in Sec.\,\ref{sec:WEL}. The last section summerizes
the main results of the paper.

\section{Classical particle-field systems\label{sec:CPFS}}

The classical non-relativistic particle-field system in flat space
is governed by the Newton-Maxwell equations 
\begin{gather}
\begin{gathered}\ddot{\boldsymbol{X}}_{sp}=\left(\frac{q}{m}\right)_{s}\left(\boldsymbol{E}+\frac{1}{c}\dot{\boldsymbol{X}}_{sp}\times\boldsymbol{B}\right),\end{gathered}
\label{eq:N}\\
\nabla\cdot\boldsymbol{E}=4\pi\sum_{s,p}q_{s}\delta(\boldsymbol{X}_{sp}-\boldsymbol{x}),\,\\
\nabla\times\boldsymbol{B}=\frac{4\pi}{c}\sum_{s,p}q_{s}\dot{\boldsymbol{X}}_{sp}\delta(\boldsymbol{X}_{sp}-\boldsymbol{x})+\frac{1}{c}\frac{\partial\boldsymbol{E}}{\partial t},\label{eq:2}\\
\nabla\times\boldsymbol{E}=-\frac{1}{c}\frac{\partial\boldsymbol{B}}{\partial t},\\
\nabla\cdot\boldsymbol{B}=0,\label{eq:4}
\end{gather}
where $\boldsymbol{X}_{sp}(t)$ as a function of time is the trajectory
of the $p$-th particle of the $s$-species, and $q_{s}$ and $m_{s}$
are the particle charge and mass, respectively. The electric field
$\boldsymbol{E}(\boldsymbol{x},t)$ and the magnetic field $\boldsymbol{B}(\boldsymbol{x},t)$
are functions of space-time. Equations \eqref{eq:N}-\eqref{eq:2}
can be expressed equivalently in the form of the Klimontovich-Maxwell
(KM) equations \cite{Davidson01-23} 
\begin{gather}
\frac{\partial F_{s}}{\partial t}+\boldsymbol{v}\cdot\frac{\partial F_{s}}{\partial\boldsymbol{x}}+\left(\frac{q}{m}\right)_{s}\left(\boldsymbol{E}+\frac{1}{c}\boldsymbol{v}\times\boldsymbol{B}\right)\cdot\frac{\partial F_{s}}{\partial\boldsymbol{v}}=0,\label{eq:5}\\
\nabla\cdot\boldsymbol{E}=4\pi\sum_{s}q_{s}\int F_{s}d_{s}^{3}\boldsymbol{v},\\
\nabla\times\boldsymbol{B}=\frac{4\pi}{c}\sum_{s}q_{s}\int F_{s}\boldsymbol{v}d^{3}\boldsymbol{v}+\frac{1}{c}\frac{\partial\boldsymbol{E}}{\partial t},\label{eq:6}
\end{gather}
where $F_{s}(\boldsymbol{x},\boldsymbol{v},t)=\sum_{p}\delta(\boldsymbol{X}_{sp}-\boldsymbol{x})\delta(\dot{\boldsymbol{X}}_{sp}-\boldsymbol{v})$
is the Klimontovich distribution function in the phase space $(\boldsymbol{x},\boldsymbol{v})$. 

Reduced models are often used in plasma physics. For example, the
electrostatic Klimontovich-Poisson (KP) system is given by 
\begin{gather}
\frac{\partial F_{s}}{\partial t}+\boldsymbol{v}\cdot\frac{\partial F_{s}}{\partial\boldsymbol{x}}+\left(\frac{q}{m}\right)_{s}\left(-\nabla\phi+\frac{1}{c}\boldsymbol{v}\times\boldsymbol{B}_{0}\right)\cdot\frac{\partial F_{s}}{\partial\boldsymbol{v}}=0,\label{eq:8}\\
\nabla^{2}\phi=-4\pi\sum_{s}q_{s}\int F_{s}d^{3}\boldsymbol{v},\label{eq:9}
\end{gather}
where $\boldsymbol{B}_{0}(\boldsymbol{x})$ is a background magnetic
field produced by steady external currents, and $\boldsymbol{E}=-\nabla\phi$
is the longitudinal electric field. Another well-known reduced model
is the Klimontovich-Darwin (KD) system \cite{Kaufman71,Jackson99-596,Qin01-477,Krause07},
\begin{gather}
\frac{\partial F_{s}}{\partial t}+\boldsymbol{v}\cdot\frac{\partial F_{s}}{\partial\boldsymbol{x}}+\left(\frac{q}{m}\right)_{s}\left(\boldsymbol{E}+\frac{1}{c}\boldsymbol{v}\times\boldsymbol{B}\right)\cdot\frac{\partial F_{s}}{\partial\boldsymbol{v}}=0,\label{eq:5-1}\\
\nabla^{2}\phi+\nabla\cdot\left(\frac{1}{c}\frac{\partial\boldsymbol{A}}{\partial t}\right)=-4\pi\sum_{s}q_{s}\int F_{s}d^{3}\boldsymbol{v},\,\,\,\\
\nabla\times(\nabla\times\boldsymbol{A})+\frac{1}{c}\frac{\partial\nabla\phi}{\partial t}=\frac{4\pi}{c}\sum_{s}q_{s}\int F_{s}\boldsymbol{v}d^{3}\boldsymbol{v},\\
\boldsymbol{E}\equiv-\frac{1}{c}\frac{\partial\boldsymbol{A}}{\partial t}-\nabla\phi,\thinspace\thinspace\thinspace\boldsymbol{B}\equiv\nabla\times\boldsymbol{A}.
\end{gather}

The local energy-momentum conservation laws for the Klimontovich-Maxwell
system \eqref{eq:5}-\eqref{eq:6} is well-known \cite{Landau75},
\begin{gather}
\frac{\partial}{\partial t}\left[\frac{\boldsymbol{E}^{2}+\boldsymbol{B}^{2}}{8\pi}+\sum_{s,p}\frac{m_{s}\dot{\boldsymbol{X}}_{sp}^{2}}{2}\delta_{2}\right]+\nabla\cdot\left[\frac{c\boldsymbol{E}\times\boldsymbol{B}}{4\pi}+\sum_{s,p}\frac{m_{s}\dot{\boldsymbol{X}}_{sp}^{2}}{2}\dot{\boldsymbol{X}}_{sp}\delta_{2}\right]=0,\label{KM-EC}\\
\frac{\partial}{\partial t}\left[\frac{\boldsymbol{E}\times\boldsymbol{B}}{4\pi c}+\sum_{s,p}m_{s}\dot{\boldsymbol{X}}_{sp}\delta_{2}\right]+\nabla\cdot\left[\frac{\boldsymbol{E}^{2}+\boldsymbol{B}^{2}}{8\pi}I-\frac{\boldsymbol{EE}+\boldsymbol{BB}}{4\pi}+\sum_{s,p}m_{s}\dot{\boldsymbol{X}}_{sp}\dot{\boldsymbol{X}}_{sp}\delta_{2}\right]=0,\label{KM-M}
\end{gather}
where we have introduced $\delta_{2}\equiv\delta(\boldsymbol{X}_{sp}-\boldsymbol{x})$
to simplify the notation. Through the following identities, 
\begin{align}
\sum_{p}\frac{\dot{\boldsymbol{X}}_{sp}^{2}}{2}\delta_{2}=\int d^{3}\boldsymbol{v}F_{s}\frac{\boldsymbol{v}^{2}}{2},\,\,\, & \sum_{p}\frac{\dot{\boldsymbol{X}}_{sp}^{2}}{2}\dot{\boldsymbol{X}}_{sp}\delta_{2}=\int d^{3}\boldsymbol{v}F_{s}\frac{\boldsymbol{v}^{2}}{2}\boldsymbol{v},\\
\sum_{p}\dot{\boldsymbol{X}}_{sp}\delta_{2}=\int d^{3}\boldsymbol{v}F_{s}\boldsymbol{v},\,\,\, & \sum_{p}\dot{\boldsymbol{X}}_{sp}\dot{\boldsymbol{X}}_{sp}\delta_{2}=\int d^{3}\boldsymbol{v}F_{s}\boldsymbol{vv},
\end{align}
the conservation laws can be expressed equivalently in terms of the
distribution function $F_{s}$,
\begin{gather}
\frac{\partial}{\partial t}\left[\frac{\boldsymbol{E}^{2}+\boldsymbol{B}^{2}}{8\pi}+\sum_{s}\int d^{3}\boldsymbol{v}F_{s}m_{s}\frac{\boldsymbol{v}^{2}}{2}\right]+\nabla\cdot\left[\frac{c\boldsymbol{E}\times\boldsymbol{B}}{4\pi}+\sum_{s}\int d^{3}\boldsymbol{v}F_{s}m_{s}\frac{\boldsymbol{v}^{2}}{2}\boldsymbol{v}\right]=0,\label{KM-EC-1}\\
\frac{\partial}{\partial t}\left[\frac{\boldsymbol{E}\times\boldsymbol{B}}{4\pi c}+\sum_{s}\int d^{3}\boldsymbol{v}F_{s}m_{s}\boldsymbol{v}\right]+\nabla\cdot\left[\frac{\boldsymbol{E}^{2}+\boldsymbol{B}^{2}}{8\pi}I-\frac{\boldsymbol{EE}+\boldsymbol{BB}}{4\pi}+\sum_{s}\int d^{3}\boldsymbol{v}F_{s}m_{s}\boldsymbol{vv}\right]=0,\label{KM-M-1}
\end{gather}

For the reduced systems, e.g., the KP system and the KD system, it
is also critical to know the exact local energy-momentum conservation
laws admitted by the models. In practical applications, such as current
drive and heating with lower-hybrid waves \cite{Fisch87}, and electrostatic
drift-wave turbulence, such local energy-momentum conservation laws
for the reduced system have profound implications \cite{Scott10,Brizard11,Sugama13}.
We emphasize that we are looking for the exact conservation laws admitted
by the KP and KD systems, which are not exact special cases of the
KM system, and should be viewed as independent systems in their own
right. For example, we cannot take the exact energy-momentum equations
\eqref{KM-EC} and \eqref{KM-M}, and approximate $\boldsymbol{E}$
by $-\nabla\phi$ and $\boldsymbol{B}$ by $\boldsymbol{B}_{0}$ to
obtain the exact energy-momentum conservation law for the KP system,
even though the conservation law obtained this way could be an approximate
one for the KP system. The existence of exact local conservation laws
is a necessary condition for the models to be theoretically well-posed
and for the validity of particle simulations based on the KP or KD
systems \cite{Qin01-477}. 

On the other hand, conservation laws and symmetries are closely related.
It is commonly believed that, according to Noether's theorem \cite{Noether1918,Olver93-242},
conservation laws can be derived from the symmetries of the corresponding
field theories. In standard field theories, this is certainly true,
and the symmetry in time for the action is related to energy conservation,
and the symmetry in space corresponds to momentum conservation. Therefore,
it is reasonable to expect that by analyzing the symmetries of the
actions and Lagrangian densities for the reduced systems considered
here, we may be able to systematically derive the desired conservation
laws. However, it is surprising to find out that for particle-field
systems considered here, the field theory works differently. First,
let's recall the action and Lagrangian density for the KM system given
by Low \cite{Low58}, 
\begin{gather}
\mathfrak{\mathcal{A}}[\phi,\boldsymbol{A},\boldsymbol{X}_{sp}]=\int L_{KM}d^{3}\boldsymbol{x}dt,\,\, L_{KM}=L_{KMF}+L_{KMP},\\
L_{KMF}=\left(\frac{1}{c}\frac{\partial\boldsymbol{A}}{\partial t}+\nabla\phi\right)^{2}/8\pi-\left(\nabla\times\boldsymbol{A}\right)^{2}/8\pi,\\
L_{KMP}=\sum_{s,p}\left[-q_{s}\phi+\frac{q_{s}}{c}\dot{\boldsymbol{X}}_{sp}\cdot\boldsymbol{A}+\frac{m_{s}}{2}\dot{\boldsymbol{X}}_{sp}^{2}\right]\delta_{2}.
\end{gather}
It is straightforward to verify that Eqs.\,\eqref{eq:N} -\eqref{eq:4}
follow from $\delta\mathfrak{\mathcal{A}}/\delta\boldsymbol{X}_{sp}=0$,
$\delta\mathfrak{\mathcal{A}}/\delta\phi=0$, and $\delta\mathfrak{\mathcal{A}}/\delta\boldsymbol{A}=0$.
For the KP system, the action and Lagrangian density are given by
\begin{gather}
\mathfrak{\mathcal{A}}[\phi,\boldsymbol{X}_{sp}]=\int L_{KP}d^{3}\boldsymbol{x}dt,\,\, L_{KP}=L_{KPF}+L_{KPP},\label{eq:19}\\
L_{KPF}=(\nabla\phi)^{2}/8\pi,\,\,\, L_{KPP}=\sum_{s,p}\left[-q_{s}\phi+\frac{q_{s}}{c}\dot{\boldsymbol{X}}_{sp}\cdot\boldsymbol{A}_{0}+\frac{m_{s}}{2}\dot{\boldsymbol{X}}_{sp}^{2}\right]\delta_{2},
\end{gather}
where $\boldsymbol{A}_{0}$ is the vector potential for a given external
magnetic field $\boldsymbol{B}_{0}=\nabla\times\boldsymbol{A}_{0}$.
For the KD system, the action and Lagrangian density for the KM system
are 
\begin{gather}
\mathfrak{\mathcal{A}}[\phi,\boldsymbol{A},\boldsymbol{X}_{sp}]=\int L_{KD}d^{3}\boldsymbol{x}dt,\,\, L_{KD}=L_{KDF}+L_{KDP},\\
L_{KDF}=\left[\frac{2}{c}\nabla\phi\cdot\frac{\partial\boldsymbol{A}}{\partial t}+(\nabla\phi)^{2}\right]/8\pi-\left(\nabla\times\boldsymbol{A}\right)^{2}/8\pi,\\
L_{KDP}=\sum_{p=1}^{N}\left[-q_{s}\phi+\frac{q_{s}}{c}\dot{\boldsymbol{X}}_{sp}\cdot\boldsymbol{A}+\frac{m_{s}}{2}\dot{\boldsymbol{X}}_{sp}^{2}\right]\delta_{2}.
\end{gather}
Based on the spirit of Noether's theorem, we would like to determine
whether the local energy-momentum conservation laws can be derived
from the symmetries of the corresponding Lagrangian density. It turns
out that the answer to this question is not as simple as that in standard
field theory. This is because the fields in the present field theory,
i.e., $\boldsymbol{X}_{sp}(t)$, $\phi(\boldsymbol{x},t)$, and $\boldsymbol{A}(\boldsymbol{x},t)$
are defined on different domains. The potentials are defined on the
space-time domain $(\boldsymbol{x},t),$ whereas the particle trajectory
$\boldsymbol{X}_{sp}(t)$ is only defined on the time-axis. This unique
feature has not been discussed before, and it makes a significant
difference in the formulation of the field theory presented here. 

In the next section, we develop the field theory for classical particle-systems
with this feature, in particular, the KM system, the KP system, and
the KD system. The most distinct characteristic of the field theory
presented here is that the field equation for $\boldsymbol{X}_{sp}(t)$
assumes a form we call the weak Euler-Lagrange (EL) equation, which
is different from the standard Euler-Lagrange equation. The necessity
of using the weak EL equation is mandated by the fact that $\boldsymbol{X}_{sp}(t),$
as a field, is not defined on the entire space-time domain, but only
on the time-axis. The weak EL equation with respect to $\boldsymbol{X}_{sp}(t)$
plays an indispensable role in the symmetry analysis and derivation
of local conservation laws. For the KP system and KD system, the analysis
developed here enables us to determine the desired local conservation
laws, which have not been systematically discussed in the literature.
For the KM system, where the local energy-momentum conservation laws
\eqref{KM-EC}-\eqref{KM-M} are known, the present analysis serves
the purpose of establishing a connection between the energy-momentum
conservation laws and symmetries of the Lagrangian density $L_{KM}.$
Interestingly, such a connection has only been cautiously suggested
\cite{Landau75} but not explicitly established previously. This is
perhaps not surprising, because the weak EL equation developed here
is needed to establish the connection. Due to the space limitation,
we present in Sec.\,\ref{sec:WEL} the detailed derivation of the
field theory and new conservation laws only for the KP system of a
magnetized plasma, and summarize the main results for the KM and the
KD systems at the end.

In plasma physics, one often works with the Vlasov-Maxwell (VM) system.
Equations \eqref{eq:5}-\eqref{eq:6} recover the VM equations when
two-particle correlations (collisions) become negligibly small as
the number of particles becomes increasingly large, while holding
total charge and and charge to mass ratio fixed. In the present study,
we work with the Klimontovich-Maxwell system \eqref{eq:5}-\eqref{eq:6}
or \eqref{eq:N}-\eqref{eq:2} and pass to the limit of the Vlasov-Maxwell
system when necessary under the assumption of negligible collisions.
Similarly, the Vlasov-Poisson (VP) and Vlasov-Darwin (VD) systems
are regarded as the collisionless limits of the KP and KD systems,
respectively. As a reduced system, the KP (or VP) system describes
many important physical processes when the characteristic velocity
of the particles or waves are much slower than the speed of light.
These include electrostatic waves in plasmas (Langmuir waves) \cite{Tonks29},
and collective dynamics and excitations in charged particle beams
in a frame moving with the beam \cite{Davidson01-23}. The fundamental
theory of Landau damping \cite{Landau46} was first developed for
the VP (or KP) system. In astrophysics, the VP (or KP) system has
also been used to model the collective dynamics of self-gravitating
systems with an attractive Newtonian potential \cite{Rein97,Andreasson05}.
Because of these important applications, the VP (or KP) system and
its associated Landau damping have also been studied with great interest
in the mathematical physics community \cite{Mouhot11,Villani14}.
We also note that while our focus here is on particle-field systems,
Eulerian field theories for the VM and VP systems have been developed
by Morrison et al. \cite{Morrison80,Ye92,Brizard00,Squire13} using
a variety of theoretical constructions. In Eulerian theories, the
particle distribution in phase space replaces $\boldsymbol{X}_{sp}(t)$
as the field variable.

\section{Weak Euler-Lagrange equation, symmetry, and conservation laws\label{sec:WEL}}

We begin with Eq.\,\eqref{eq:19} for the KP system, and determine
how the action and Lagrangian density vary in response to the field
variation $\delta\boldsymbol{X}_{sp}$ and $\delta\phi(\boldsymbol{x},t)$,
\begin{gather}
\mathfrak{\delta\mathcal{A}}=\int d^{3}\boldsymbol{x}dt\,\delta\phi E_{\phi}(L_{KP})+\sum_{s,p=1}^{N}\int dt\,\delta\boldsymbol{X}_{sp}\cdot\int d^{3}\boldsymbol{x}\, E_{\boldsymbol{X}_{sp}}(L_{KP}),\label{eq:24}\\
E_{\phi}(L_{KP})\equiv\frac{\partial L_{KP}}{\partial\phi}-\frac{D}{Dx^{i}}\frac{\partial L_{KP}}{\partial\phi_{,i}},\,\,\, E_{\boldsymbol{X}_{sp}}(L_{KP})\equiv\frac{\partial L_{KP}}{\partial\boldsymbol{X}_{sp}}-\frac{D}{Dt}\left(\frac{\partial L_{KP}}{\partial\dot{\boldsymbol{X}}_{sp}}\right).
\end{gather}
In Eq.\,\eqref{eq:24}, $\phi_{,i}\equiv\partial\phi/\partial x^{i}$
and integration by parts has been applied with respect to terms containing
$\partial L_{KP}/\partial\phi_{,i}$ and $\partial L_{KP}/\partial\dot{\boldsymbol{X}}_{sp}$.
Here, $E_{\phi}(L_{KP})$ and $E_{\boldsymbol{X}_{sp}}(L_{KP})$ are
the Euler operators with respect to $\phi$ and $\boldsymbol{X}_{sp}$,
respectively. For a variable $h,$ $Dh/Dx^{i}$ and $Dh/Dt$ represent
the space-time derivatives when $h=h(\boldsymbol{x},t)$ is considered
as a field on the space-time domain. Because $\delta\phi(\boldsymbol{x},t)$
is arbitrary, $\mathfrak{\delta\mathcal{A}}/\delta\phi=0$ requires
the Euler-Lagrange (EL) equation for $\phi$ to hold, i.e., $E_{\phi}(L_{KP})=0$,
which is indeed the Poisson equation \eqref{eq:9}, as expected. The
field equation for $\boldsymbol{X}_{sp}$ is more interesting. Because
$\delta\boldsymbol{X}_{sp}$ is arbitrary only on the time-axis, the
condition $\mathfrak{\delta\mathcal{A}}/\delta\boldsymbol{X}_{sp}=0$
requires only that the integral of $E_{\boldsymbol{X}_{sp}}(L_{KP})$
over space vanish, i.e., 
\begin{gather}
\int d^{3}\boldsymbol{x}E_{\boldsymbol{X}_{sp}}(L_{KP})=0.\label{eq:26}
\end{gather}
Equation \eqref{eq:26} will be called the submanifold Euler-Lagrangian
equation because it is defined only on the time-axis after the integrating
over the spatial variable. If $\boldsymbol{X}_{sp}$ were a function
of the entire space-time domain, then $E_{\boldsymbol{X}_{sp}}(L_{KP})$
would vanish everywhere, as in the case for $\phi(\boldsymbol{x},t)$.
In general, we expect that $E_{\boldsymbol{X}_{sp}}(L_{KP})\neq0$.

We now derive an explicit expression for $E_{\boldsymbol{X}_{sp}}(L_{KP})$.
For the first term in $E_{\boldsymbol{X}_{sp}}(L_{KP})$, 
\begin{align}
\frac{\partial L_{KP}}{\partial\boldsymbol{X}_{sp}} & =\left(\frac{q_{s}}{c}\boldsymbol{A}_{0}\cdot\dot{\boldsymbol{X}}{}_{sp}-q_{s}\phi+\frac{m_{s}}{2}\dot{\boldsymbol{X}}_{sp}^{2}\right)\frac{\partial\delta_{2}}{\partial\boldsymbol{X}_{sp}}\nonumber \\
 & =\frac{\partial}{\partial\boldsymbol{x}}\left(H_{sp}-\dot{\boldsymbol{X}}{}_{sp}\cdot\boldsymbol{P}_{sp}\right)+\left(\frac{q_{s}}{c}\frac{\partial\boldsymbol{A}_{0}}{\partial\boldsymbol{x}}\cdot\dot{\boldsymbol{X}}{}_{sp}-q_{s}\frac{\partial\phi}{\partial\boldsymbol{x}}\right)\delta_{2},\label{eq:27}
\end{align}
where the momentum $\boldsymbol{P}_{sp}$ density and Hamiltonian
$H_{sp}$ density are defined as
\begin{equation}
\boldsymbol{P}_{sp}(\boldsymbol{x},t)\equiv\frac{\partial L_{KP}}{\partial\dot{\boldsymbol{X}}_{sp}}=\left(m_{s}\dot{\boldsymbol{X}}{}_{sp}+\frac{q_{s}}{c}\boldsymbol{A}_{0}\right)\delta_{2},\thinspace\thinspace\thinspace H_{sp}(\boldsymbol{x},t)\equiv\left(q_{s}\phi+\frac{m_{s}}{2}\dot{\boldsymbol{X}}{}_{sp}^{2}\right)\delta_{2}.\label{eq:P}
\end{equation}
The second term in $E_{\boldsymbol{X}_{sp}}(L_{KP})$ is given by
\begin{align}
\frac{D}{Dt}\frac{\partial L_{KP}}{\partial\dot{\boldsymbol{X}}_{sp}} & =m_{s}\ddot{\boldsymbol{X}}_{sp}\delta_{2}+\left(m_{s}\dot{\boldsymbol{X}}{}_{sp}+\frac{q_{s}}{c}\boldsymbol{A}_{0}\right)\frac{\partial\delta_{2}}{\partial t}\nonumber \\
 & =m_{s}\ddot{\boldsymbol{X}}_{sp}\delta_{2}-\frac{\partial}{\partial\boldsymbol{x}}\cdot\left(\dot{\boldsymbol{X}}{}_{sp}\boldsymbol{P}_{sp}\right)+\frac{q_{s}}{c}\dot{\boldsymbol{X}}{}_{sp}\cdot\frac{\partial\boldsymbol{A}_{0}}{\partial\boldsymbol{x}}\delta_{2}.\label{eq:21}
\end{align}
Therefore, 
\begin{align}
E_{\boldsymbol{X}_{sp}}(L_{KP}) & =\left[\frac{q_{s}}{c}\left(\frac{\partial\boldsymbol{A}_{0}}{\partial\boldsymbol{x}}\cdot\dot{\boldsymbol{X}}{}_{sp}-\dot{\boldsymbol{X}}{}_{sp}\cdot\frac{\partial\boldsymbol{A}_{0}}{\partial\boldsymbol{x}}\right)-q_{s}\frac{\partial\phi}{\partial\boldsymbol{x}}-m_{s}\ddot{\boldsymbol{X}}_{sp}\right]\delta_{2}\nonumber \\
 & +\frac{\partial}{\partial\boldsymbol{x}}\left(H_{sp}-\dot{\boldsymbol{X}}{}_{sp}\cdot\boldsymbol{P}_{sp}\right)+\frac{\partial}{\partial\boldsymbol{x}}\cdot\left(\dot{\boldsymbol{X}}{}_{sp}\boldsymbol{P}_{sp}\right).\label{eq:Exp}
\end{align}
Substituting Eq.\,\eqref{eq:Exp} into the submanifold EL equation
\eqref{eq:26}, we immediately recover Newton's equation for $\boldsymbol{X}_{sp}$,
i.e., 
\begin{equation}
\frac{m_{s}}{q_{s}}\ddot{\boldsymbol{X}}=-\frac{\partial\phi}{\partial\boldsymbol{x}}+\frac{1}{c}\dot{\boldsymbol{X}}{}_{sp}\times\boldsymbol{B}_{0},
\end{equation}
 which reduces Eq.\,\eqref{eq:Exp} to 
\begin{align}
E_{\boldsymbol{X}_{sp}}(L_{KP}) & \equiv\frac{\partial L_{KP}}{\partial\boldsymbol{X}_{sp}}-\frac{D}{Dt}\left(\frac{\partial L_{KP}}{\partial\dot{\boldsymbol{X}}_{sp}}\right)=\frac{\partial}{\partial\boldsymbol{x}}\left(H_{sp}-\dot{\boldsymbol{X}}{}_{sp}\cdot\boldsymbol{P}_{sp}\right)+\frac{\partial}{\partial\boldsymbol{x}}\cdot\left(\dot{\boldsymbol{X}}{}_{sp}\boldsymbol{P}_{sp}\right).\label{eq:WEL}
\end{align}
As expected, $E_{\boldsymbol{X}_{sp}}(L_{KP})\neq0$ . We will refer
to Eq.\,\eqref{eq:WEL} as the weak Euler-Lagrange equation, which
is the foundation for the subsequent analysis of the local conservation
laws. The qualifier ``weak'' is used to indicate the fact that only
the spatial integral of the Euler derivative $E_{\boldsymbol{X}_{sp}}(L_{KP})$
is zero {[}see Eq.\,\eqref{eq:26}{]}, in comparison with the standard
EL equation, which demands that the Euler derivative vanishes everywhere.

We define a symmetry of the action $\mathfrak{\mathcal{A}}[\phi,\boldsymbol{X}_{sp}]$
to be a group of transformation 
\begin{equation}
(\boldsymbol{x},t,\phi,\boldsymbol{X}_{sp})\mapsto(\tilde{\boldsymbol{x}},\tilde{t},\tilde{\phi},\tilde{\boldsymbol{X}}_{sp})
\end{equation}
such that 
\begin{equation}
\int L_{KP}[\boldsymbol{x},t,\phi,\boldsymbol{X}_{sp}]d^{3}\boldsymbol{x}dt=\int L_{KP}[\tilde{\boldsymbol{x}},\tilde{t},\tilde{\phi},\tilde{\boldsymbol{X}}_{sp}]d^{3}\tilde{\boldsymbol{x}}d\tilde{t}.
\end{equation}
If the symmetry is generated by a vector field on the space of $(\boldsymbol{x},t,\phi,\boldsymbol{X}_{sp}),$
\[
V=\boldsymbol{\xi}\cdot\frac{\partial}{\partial\boldsymbol{x}}+\xi^{t}\frac{\partial}{\partial t}+\psi\frac{\partial}{\partial\phi}+\boldsymbol{Y}_{p}\cdot\frac{\partial}{\partial\boldsymbol{X}_{sp}},
\]
 then the infinitesimal criteria of invariance is given by \cite{Olver93-242}
\begin{equation}
prV(L)+L\, Div\xi=0,\label{eq:isc}
\end{equation}
where $Div\xi$ is the divergence of the vector field 
\begin{equation}
\xi=\boldsymbol{\xi}\cdot\frac{\partial}{\partial\boldsymbol{x}}+\xi^{t}\frac{\partial}{\partial t}
\end{equation}
 on the space-time domain, and $prV$ is the prolongation of the vector
field $V$ on $(\boldsymbol{x},t,\phi,\boldsymbol{X}_{sp}).$ The
prolongation $prV$ is a vector field on the jet space, consisting
of the space of $(\boldsymbol{x},t,\phi,\boldsymbol{X}_{sp})$ and
the space of derivatives of $(\phi,\boldsymbol{X}_{sp})$ with respect
to $(\boldsymbol{x},t)$. A comprehensive description of this subject
can be found in Ref. \cite{Olver93-242}. Given the symmetry vector
field $V$, the infinitesimal criteria for invariance will generate
the desired conservation law corresponding to the symmetry vector
field $V,$ after use is made of the EL equation as well as the weak
EL equation for the systems in the present study. We first look for
the symmetry group that generates local energy conservation. The group
of transformation 
\begin{equation}
(\tilde{\boldsymbol{x}},\tilde{t},\tilde{\phi},\tilde{\boldsymbol{X}_{sp}})=(\boldsymbol{x},t+\epsilon,\phi,\boldsymbol{X}_{sp}),\thinspace\thinspace\thinspace\thinspace\epsilon\in R\label{eq:tt}
\end{equation}
is a symmetry of $L_{KP},$ because $L_{KP}$ does not depend on $t$
explicitly, i.e., $\partial L_{KP}/\partial t=0$, which can be written
as 
\begin{equation}
\frac{DL_{KP}}{Dt}-\phi_{,t}\frac{\partial L_{KP}}{\partial\phi}-\phi_{,jt}\frac{\partial L_{KP}}{\partial\phi_{,j}}-\sum_{s,p}\left(\dot{\boldsymbol{X}}{}_{sp}\cdot\frac{\partial L_{KP}}{\partial\boldsymbol{X}_{sp}}+\ddot{\boldsymbol{X}}_{sp}\cdot\frac{\partial L_{KP}}{\partial\dot{\boldsymbol{X}}_{sp}}\right)=0.\label{eq:33-1}
\end{equation}
Equation \eqref{eq:33-1} is the special form of Eq.\,\eqref{eq:isc}
for this symmetry group. From the EL equation for $\phi$, i.e., $E_{\phi}(L_{KP})=0$,
we obtain 
\begin{equation}
\phi_{,t}\frac{\partial L_{KP}}{\partial\phi}+\phi_{,jt}\frac{\partial L_{KP}}{\partial\phi_{,j}}=\frac{D}{Dx^{j}}\left(\phi_{,t}\frac{\partial L_{KP}}{\partial\phi_{,j}}\right).\label{eq:Ephi2}
\end{equation}
The weak EL equation for $\boldsymbol{X}_{sp}$, i.e., Eq.\,\eqref{eq:WEL},
gives 
\begin{align}
\dot{\boldsymbol{X}}{}_{sp}\cdot\frac{\partial L_{KP}}{\partial\boldsymbol{X}_{sp}}+\ddot{\boldsymbol{X}}_{sp}\cdot\frac{\partial L_{KP}}{\partial\dot{\boldsymbol{X}}{}_{sp}} & =\frac{\partial}{\partial\boldsymbol{x}}\cdot\left[\dot{\boldsymbol{X}}_{sp}\left(q_{s}\phi+\frac{m_{s}}{2}\dot{\boldsymbol{X}}{}_{sp}^{2}\right)\delta_{2}\right]+\frac{D}{Dt}\left(\dot{\boldsymbol{X}}{}_{sp}\cdot\boldsymbol{P}_{sp}\right)\label{eq:Exp2}
\end{align}
Combining Eqs.\,\eqref{eq:Ephi2} and \eqref{eq:Exp2}, we obtain
the first local energy conservation law, 
\begin{equation}
\frac{\partial}{\partial t}\left[\frac{(\nabla\phi)^{2}}{8\pi}-\sum_{s,p}\left(q_{s}\phi+\frac{m_{s}}{2}\dot{\boldsymbol{X}}^{2}{}_{sp}\right)\delta_{2}\right]+\frac{\partial}{\partial\boldsymbol{x}}\cdot\left[\frac{-1}{4\pi}\phi_{,t}\nabla\phi-\sum_{s,p}\dot{\boldsymbol{X}}_{sp}\left(q_{s}\phi+\frac{m_{s}}{2}\dot{\boldsymbol{X}}{}_{sp}^{2}\right)\delta_{2}\right]=0.\label{eq:33}
\end{equation}
We subtract the identify 
\begin{equation}
\frac{1}{4\pi}\frac{\partial}{\partial t}\left[(\nabla\phi)^{2}+\phi\nabla^{2}\phi\right]+\frac{1}{4\pi}\frac{\partial}{\partial\boldsymbol{x}}\cdot\left(-\phi_{,t}\nabla\phi-\phi\nabla\phi_{,t}\right)=0
\end{equation}
from Eq.\,\eqref{eq:33} to express the energy conservation law in
another equivalent form 
\begin{equation}
\frac{\partial}{\partial t}\left[\frac{(\nabla\phi)^{2}}{8\pi}+\sum_{s,p}\frac{m_{s}\dot{\boldsymbol{X}}{}_{sp}^{2}}{2}\delta_{2}\right]+\frac{\partial}{\partial\boldsymbol{x}}\cdot\left[\sum_{s,p}\boldsymbol{\dot{X}}_{sp}\left(q_{s}\phi+\frac{m_{s}\dot{\boldsymbol{X}}{}_{sp}^{2}}{2}\right)\delta_{2}-\frac{1}{4\pi}\phi\nabla\phi_{,t}\right]=0.\label{eq:37}
\end{equation}
In terms of the distribution function $F_{s},$ we obtain 
\begin{align}
 & \frac{\partial}{\partial t}\left[\frac{(\nabla\phi)^{2}}{8\pi}+\sum_{s}\int F_{s}\frac{m_{s}\boldsymbol{v}^{2}}{2}d^{3}\boldsymbol{v}\right]+\nonumber \\
 & \frac{\partial}{\partial\boldsymbol{x}}\cdot\left(\sum_{s}\int F_{s}\frac{m_{s}\boldsymbol{v}^{2}}{2}\boldsymbol{v}d^{3}\boldsymbol{v}+\sum_{s}q_{s}\phi\int F_{s}\boldsymbol{v}d^{3}\boldsymbol{v}-\frac{1}{4\pi}\phi\nabla\phi_{,t}\right)=0.\label{KP-EC}
\end{align}

We emphasize again that Eq.\,\eqref{KP-EC} is the exact energy conservation
law admitted by the KP system Eqs.\,\eqref{eq:8} and \eqref{eq:9},
and it cannot be obtained by replacing $\boldsymbol{E}$ by $-\nabla\phi$
and $\boldsymbol{B}$ by $\boldsymbol{B}_{0}$ in the energy conservation
law for the KM sytem\,\eqref{KM-EC}. The sum of the last two terms
in Eq.\,\eqref{KP-EC} is the electrostatic Poynting flux of the
KP system, first discussed by Similon \cite{Similon81} for an unmagnetized
plasma by algebraic manipulation. Its importance for electrostatic
particle simulations was addressed by Decyk \cite{Decyk82}. Here,
it appears naturally as a consequence of the symmetry analysis. We
observe that the external $\boldsymbol{B}_{0}$ does not contribute
to the energy flux of the electromagnetic field. To further appreciate
the importance of the exact energy conservation law \eqref{KP-EC},
let's consider the well-established technique of current drive and
heating of a magnetized plasma using electrostatic lower-hybrid (LH)
waves \cite{Fisch87}, which are adequately described by the KM (or
KP) system. In this application, the energy and momentum of the LH
waves are converted to those of the particles, and it is of practical
importance to know the heating power of a specific LH wave system.
However, if we calculated the energy flux of the LH waves from Eq.\,\eqref{KM-EC}
by replacing $\boldsymbol{E}$ by $-\nabla\phi$ and $\boldsymbol{B}$
by $\boldsymbol{B}_{0}$, we would find that $\nabla\cdot[\nabla\phi\times\boldsymbol{B}_{0}]=0,$
i.e., the LH waves do not carry an energy flux. This is obviously
erroneous. The typical power of such systems in modern magnetic fusion
devices is several megawatts. The correct way to calculate the energy
flux of the LH waves is to use Eq.\,\eqref{KP-EC} instead. Specifically,
the several megawatts of energy carried by the LH waves flow into
the plasma through the last two terms in Eq.\,\eqref{KP-EC}. 

Up to now, we have treated the KM and KP systems as independent systems,
each of which has its own governing equations, Lagrangian, and conservation
laws. On the other hand, it is also correct to treat the KP system
as the electrostatic approximation to the KM system. From the perspective
of the governing equations, this approximation is equivalent to replacing
$\boldsymbol{E}$ by $-\nabla\phi$ and $\boldsymbol{B}$ by $\boldsymbol{B}_{0}$
in the KM system. But this simple procedure does not work for the
corresponding conservation laws. What is needed here is a more rigorous
procedure to derive the electrostatic approximation that reduces from
the KM system to the KP system. After this rigorous procedure is carried
out, we find that the correct energy conservation law for the KP system
obtained from that of the KM system is actually Eq.\,\eqref{KP-EC},
instead of that obtained from Eq.\,\eqref{KM-EC} by replacing $\boldsymbol{E}$
by $-\nabla\phi$ and $\boldsymbol{B}$ by $\boldsymbol{B}_{0}$.
A similar argument applies to the momentum conservation law for the
KP system, i.e., Eq.\,\eqref{eq:KPmc2}. These derivations are given
in detail in the Appendix. 

Our next goal is to search for the symmetry that generates the momentum
conservation law. In standard field theories, if the Lagrangian density
does not depend on $\boldsymbol{x}$ explicitly, then it admits the
symmetry of spatial translation, $\tilde{\boldsymbol{x}}=\boldsymbol{x}+\epsilon\boldsymbol{u}$,
for a constant vector $\boldsymbol{u}$ and $\epsilon\in R.$ Then
the usual form of Noether's theorem leads to momentum conservation.
This strategy does not work here because $L_{KP}$ depends on $\boldsymbol{x}$
explicitly through $\delta_{2}\equiv\delta(\boldsymbol{X}_{sp}-\boldsymbol{x})$
and $\boldsymbol{A}_{0}(\boldsymbol{x}).$ However, if we simultaneously
translate both $\boldsymbol{x}$ and $\boldsymbol{X}_{sp}$ by the
same amount, then $\delta_{2}$ is invariant. Thus, we consider the
translational transformation 
\begin{equation}
(\tilde{\boldsymbol{x}},\tilde{t},\tilde{\phi},\tilde{\boldsymbol{X}}_{sp})=(\boldsymbol{x}+\epsilon\boldsymbol{u},t,\phi,\boldsymbol{X}_{sp}+\epsilon\boldsymbol{u}),\thinspace\thinspace\thinspace\thinspace\epsilon\in R\label{eq:st}
\end{equation}
under which $\tilde{\phi}(\tilde{\boldsymbol{x}})=\phi(\boldsymbol{x})=\phi(\tilde{\boldsymbol{x}}-\epsilon\boldsymbol{u})$
and $\tilde{\boldsymbol{X}}_{sp}(\tilde{t})=\boldsymbol{X}_{sp}(t)+\epsilon\boldsymbol{u}.$
When $\boldsymbol{A}_{0}(\boldsymbol{x})=0$, we can verify that Eq.\,\eqref{eq:isc}
is satisfied, and Eq.\,\eqref{eq:st} is indeed a symmtry admitted
by $L_{KP}$. The corresponding vector field is 
\begin{equation}
V=\frac{\partial}{\partial\boldsymbol{x}}+\sum_{sp}\frac{\partial}{\partial\boldsymbol{X}_{sp}}
\end{equation}
, and $V$ is the only non-vanishing component of $PrV$ since it
is a constant. The notation $\partial/\partial\boldsymbol{x}$ here
represents $\partial/\partial x^{i}$ for $i=1,2,3.$ In this case,
the infinitesimal criteria of invariance in \eqref{eq:isc} is 
\begin{equation}
\frac{\partial L_{KP}}{\partial\boldsymbol{x}}+\sum_{p}\frac{\partial L}{\partial\boldsymbol{X}_{sp}}=0.\label{eq:39}
\end{equation}
When $\boldsymbol{A}_{0}(\boldsymbol{x})\neq0$, the right-hand side
of Eq.\,\eqref{eq:39} will have a source term, and instead we obtain
\begin{equation}
\frac{\partial L_{KP}}{\partial\boldsymbol{x}}+\sum_{s,p}\frac{\partial L}{\partial\boldsymbol{X}_{sp}}=\sum_{s,p}\dot{\boldsymbol{X}}{}_{sp}\cdot\frac{\partial\boldsymbol{A}_{0}}{\partial\boldsymbol{x}}\delta_{2}.\label{eq:40}
\end{equation}
It will be clear shortly that this term represents a part of the momentum
input due to the external magnetic field through the Lorentz force.
For the first term in Eq.\,\eqref{eq:40}, we invoke the EL equation
$E_{\phi}(L_{KP})=0$ to obtain 
\begin{equation}
\frac{\partial L_{KP}}{\partial\boldsymbol{x}}=\frac{DL_{KP}}{D\boldsymbol{x}}-\frac{D}{Dx^{j}}\left(\frac{\partial L_{KP}}{\partial\phi_{,j}}\nabla\phi\right).
\end{equation}
For the second term in Eq.\,\eqref{eq:40}, the weak EL equation
for $\boldsymbol{X}_{sp}$ \eqref{eq:WEL} is applied, which gives
\begin{alignat}{1}
\frac{\partial L}{\partial\boldsymbol{X}_{sp}}= & \frac{D\boldsymbol{P}_{sp}}{Dt}+\frac{\partial}{\partial\boldsymbol{x}}\left(H_{sp}-\dot{\boldsymbol{X}}{}_{sp}\cdot\boldsymbol{P}_{sp}\right)+\frac{\partial}{\partial\boldsymbol{x}}\cdot\left(\dot{\boldsymbol{X}}{}_{sp}\boldsymbol{P}_{sp}\right)\label{eq:41}
\end{alignat}
Therefore, the conservation law generated by Eq.\,\eqref{eq:40}
is 
\begin{equation}
\frac{\partial}{\partial t}\left(\sum_{s,p}m_{s}\dot{\boldsymbol{X}}{}_{sp}\delta_{2}\right)+\frac{\partial}{\partial\boldsymbol{x}}\cdot\left[\sum_{s,p}m_{s}\dot{\boldsymbol{X}}{}_{sp}\dot{\boldsymbol{X}}{}_{sp}\delta_{2}+\frac{\boldsymbol{I}}{8\pi}(\nabla\phi)^{2}-\frac{1}{4\pi}\nabla\phi\nabla\phi\right]=\sum_{s,p}m_{s}\frac{\dot{\boldsymbol{X}}{}_{sp}}{c}\times\boldsymbol{B}_{0}\delta_{2}.\label{eq:KPmc}
\end{equation}
Evidently, this is the local momentum conservation. In terms of the
distribution function $F_{s}$, it can be expressed as 
\begin{align}
\frac{\partial}{\partial t}\left(\sum_{s}m_{s}\int F_{s}\boldsymbol{v}d^{3}\boldsymbol{v}\right) & +\frac{\partial}{\partial\boldsymbol{x}}\cdot\left[\sum_{s}m_{s}\int F_{s}\boldsymbol{vv}d^{3}\boldsymbol{v}+\frac{\boldsymbol{I}}{8\pi}(\nabla\phi)^{2}-\frac{1}{4\pi}\nabla\phi\nabla\phi\right]\nonumber \\
 & =\sum_{s}q_{s}\left(\int F_{s}\frac{\boldsymbol{v}}{c}d^{3}\boldsymbol{v}\right)\times\boldsymbol{B}_{0}.\label{eq:KPmc2}
\end{align}
The first term on the left-hand side of Eq.\,\eqref{eq:KPmc} or
Eq.\,\eqref{eq:KPmc2} is the rate of variation of the momentum density,
the second term is the divergence of the flux, and the term on the
right-hand side is the momentum input due to the background magnetic
field. Note that the momentum density is purely mechanical, and does
not include the electromagnetic momentum density $-\nabla\phi\times\boldsymbol{B}_{0}/4\pi c.$
This is not totally intuitive. This conservation law is the result
of the symmetry \eqref{eq:st}, which is different from the well-known
translational symmetry for standard field theory. Because $L_{KP}$
depends on $\boldsymbol{x}$ explicitly through $\delta_{2}\equiv\delta(\boldsymbol{X}_{sp}-\boldsymbol{x})$,
a translation in $\boldsymbol{x}$ alone is not a symmetry of $L_{KP},$
even when $\boldsymbol{A}_{0}(\boldsymbol{x})=0$. Instead, the symmetry
group \eqref{eq:st} simultaneously translates the space $\boldsymbol{x}$
and the field $\boldsymbol{X}_{sp}$ by the same amount.

For the KD system, the weak EL equation for $\boldsymbol{X}_{sp}$
is
\begin{align}
E_{\boldsymbol{X}_{sp}}(L_{KD}) & \equiv\frac{\partial L_{KD}}{\partial\boldsymbol{X}_{sp}}-\frac{D}{Dt}\frac{\partial L_{KD}}{\partial\boldsymbol{\dot{X}}_{sp}}\nonumber \\
 & =\frac{\partial}{\partial\boldsymbol{x}}\left[\left(-\boldsymbol{A}\cdot\dot{\boldsymbol{X}}{}_{sp}+\phi-\frac{1}{2}\dot{\boldsymbol{X}}{}_{sp}^{2}\right)\delta_{2}\right]+\frac{\partial}{\partial\boldsymbol{x}}\cdot\left[\dot{\boldsymbol{X}}{}_{sp}\left(\dot{\boldsymbol{X}}{}_{sp}+\boldsymbol{A}\right)\delta_{2}\right].\label{eq:WEL2-1}
\end{align}
Energy conservation follows from the infinitesimal criteria \eqref{eq:isc}
for the symmetry transformation \,\eqref{eq:tt} after the weak EL
equation \eqref{eq:WEL2-1} for $\boldsymbol{X}_{sp}$ and the EL
equations for $\phi$ and $\boldsymbol{A}$ are applied, i.e., 
\[
\frac{\partial}{\partial t}\left[\frac{(\nabla\phi)^{2}+\boldsymbol{B}^{2}}{8\pi}+\sum_{s}\int F_{s}\frac{m_{s}\boldsymbol{v}^{2}}{2}d^{3}\boldsymbol{v}\right]+\frac{\partial}{\partial\boldsymbol{x}}\cdot\left(\sum_{s}\int F_{s}\frac{m_{s}\boldsymbol{v}^{2}}{2}\boldsymbol{v}d^{3}\boldsymbol{v}+\frac{\phi_{,t}\boldsymbol{A}_{,t}+\boldsymbol{E}\times\boldsymbol{B}}{4\pi}\right)=0.
\]
Similarly, the infinitesimal criteria for the symmetry group \eqref{eq:st}
gives the momentum conservation relation
\begin{align}
 & \frac{\partial}{\partial t}\left(\sum_{s}m_{s}\int F_{s}\boldsymbol{v}d^{3}\boldsymbol{v}+\frac{\boldsymbol{E}\times\boldsymbol{B}}{4\pi}\right)\nonumber \\
 & +\frac{\partial}{\partial\boldsymbol{x}}\cdot\left[\sum_{s}m_{s}\int F_{s}\boldsymbol{vv}d^{3}\boldsymbol{v}+\frac{(\nabla\phi)^{2}+\boldsymbol{B}^{2}+2\nabla\phi\cdot\boldsymbol{A}_{,t}}{8\pi}\boldsymbol{I}-\frac{\boldsymbol{EE}+\boldsymbol{BB}-\boldsymbol{A}_{,t}\boldsymbol{A}_{,t}}{4\pi}\right]=0.\label{eq:KDmc}
\end{align}

For the KM system, the weak EL equation for $\boldsymbol{X}_{sp}$
is 
\begin{align}
E_{\boldsymbol{X}_{sp}}(L_{KD}) & \equiv\frac{\partial L_{KD}}{\partial\boldsymbol{X}_{sp}}-\frac{D}{Dt}\frac{\partial L_{KD}}{\partial\boldsymbol{\dot{X}}_{sp}}\nonumber \\
 & =\frac{\partial}{\partial\boldsymbol{x}}\left[\left(-\boldsymbol{A}\cdot\dot{\boldsymbol{X}}{}_{sp}+\phi-\frac{1}{2}\dot{\boldsymbol{X}}{}_{sp}^{2}\right)\delta_{2}\right]+\frac{\partial}{\partial\boldsymbol{x}}\cdot\left[\dot{\boldsymbol{X}}{}_{sp}\left(\dot{\boldsymbol{X}}{}_{sp}+\boldsymbol{A}\right)\delta_{2}\right].\label{eq:WEL2}
\end{align}
The symmetry groups \eqref{eq:tt} and \eqref{eq:st} gives the energy
and momentum conservation laws \eqref{KM-EC} and \eqref{KM-M} after
the weak EL equation for $\boldsymbol{X}_{sp}$ and EL equations for
$\phi$ and $\boldsymbol{A}$ are applied.

\section{Summary and Conclusions}

In summary, a close examination of the field theory for classical
particle-field systems reveals that the particle field $\boldsymbol{X}_{sp}$
and the electromagnetic field reside on different manifolds. This
unique feature is fount to imply that $E_{\boldsymbol{X}_{sp}}(L)$,
the Euler derivative of the Lagrangian density $L$ with respect to
particle's trajectory $\boldsymbol{X}_{sp}$, does not vanish on the
space-time manifold, which is surprisingly different from the standard
field theory. In fact, 
\begin{equation}
E_{\boldsymbol{X}_{sp}}(L)\equiv\frac{\partial L}{\partial\boldsymbol{X}_{sp}}-\frac{D}{Dt}\left(\frac{\partial L}{\partial\dot{\boldsymbol{X}}_{sp}}\right)=\frac{\partial}{\partial\boldsymbol{x}}\cdot\boldsymbol{T},\label{eq:wELg}
\end{equation}
for some non-vanishing tensor $\boldsymbol{T}$. Equation \eqref{eq:wELg}
is what we call the weak Euler-Lagrange equation, and it is the most
essential component in establishing the connection between energy-momentum
conservation and space-time symmetry for classical particle-field
systems. In fact, the energy-momentum conservation law follows from
the infinitesimal criteria of the space-time system, after the weak
Euler-Lagrange equation is applied. The non-vanishing tensor $\boldsymbol{T}$
is a new type of flux called the weak Euler-Lagrange current that
enters the conservation laws. For the Klimontovich-Maxwell (or Vlasov-Maxwell)
system, this theoretical construction explicitly links the well-known
energy-momentum conservation law with the space-time symmetry, which
was only cautiously suggested previously. For reduced systems, such
as the Klimontovich-Poisson (or Vlasov-Poisson) system and the Klimontovich-Darwin
(Vlasov-Darwin) system, this theoretical construction enable us to
start from fundamental symmetry properties in order to systematically
derive the energy-momentum conservation laws, which are difficult
to determine otherwise.

\section*{Appendix}

In the Sec.\,\ref{sec:WEL}, we have treated the Klimontovich-Maxwell
(KM) and Klimontovich-Poisson (KP) systems as independent systems,
each of which has its own governing equations, Lagrangian, and conservation
laws. The Vlasov-Poisson (VP) and Vlasov-Darwin (VD) systems are regarded
as the collisionless limits of the KP and KD systems, respectively.
As a reduced system, the KP (or VP) system describes many important
physical processes when the characteristic velocity of the particles
or waves are much slower than the speed of light. These include electrostatic
waves in plasmas (Langmuir waves) \cite{Tonks29}, and collective
dynamics and excitations in charged particle beams in a frame moving
with the beam \cite{Davidson01-23}. The fundamental theory of Landau
damping \cite{Landau46} was first developed for the VP (or KP) system.
In astrophysics, the VP (or KP) system has also been used to model
the collective dynamics of self-gravitating systems with an attractive
Newtonian potential \cite{Rein97,Andreasson05}. Because of these
important applications, the VP (or KP) system and its associated Landau
damping have also been studied with great interest in the mathematical
physics community \cite{Mouhot11,Villani14}. 

On the other hand, it is also correct to treat the KP system as the
electrostatic approximation to the KM system. From the perspective
of the governing equations, this approximation is equivalent to replacing
$\boldsymbol{E}$ by $-\nabla\phi$ and $\boldsymbol{B}$ by $\boldsymbol{B}_{0}$
in the KM system. But this simple procedure does not work for the
corresponding conservation laws. In this section, we present a more
rigorous procedure to carry out the electrostatic approximation that
passes from the KM system to the KP system. After this rigorous procedure
is carried out, we find that the correct energy conservation law for
the KP system obtained from that of the KM system is actually Eq.\,\eqref{KM-EC},
instead of that obtained from Eq.\,\eqref{KM-EC} by replacing $\boldsymbol{E}$
by $-\nabla\phi$ and $\boldsymbol{B}$ by $\boldsymbol{B}_{0}$. 

The electrostatic approximation applies when the characteristic velocity
of the particles $v$ and phase velocity of the waves $\omega/k$
is much slower than the speed of light, i.e., when $v/c\sim\omega/ck\sim\epsilon\ll1.$
When this condition is satisfied, it turns out that the KM (or VP)
system admits solutions with the following ordering
\begin{align}
\boldsymbol{E}_{l}= & \boldsymbol{E}_{l}^{(0)}+\epsilon\boldsymbol{E}_{l}^{(1)}+\epsilon^{2}\boldsymbol{E}_{l}^{(2)}+O(\epsilon^{3}),\\
\boldsymbol{E}_{t}= & \epsilon^{2}\boldsymbol{E}_{t}^{(2)}+O(\epsilon^{3}),\\
\boldsymbol{B}= & \boldsymbol{B}_{0}+\epsilon\boldsymbol{B}^{(1)}+\epsilon^{2}\boldsymbol{B}^{(2)}+O(\epsilon^{3}),\\
F_{s}= & F_{s}^{(0)}+\epsilon F_{s}^{(1)}+\epsilon^{2}F_{s}^{(2)}+O(\epsilon^{3}),
\end{align}
where $\boldsymbol{E}_{l}$ and $\boldsymbol{E}_{t}$ are the longitudinal
and transverse components of the electric field, respectively, and
$\boldsymbol{B}_{0}$ is the externally applied magnetic field with
$\nabla\times\boldsymbol{B}_{0}=0$ inside the plasma. The superscripts
``$(0)$'', ``$(1)$'', and ``$(2)$''represent the orders $\epsilon^{0}$,
$\epsilon^{1}$, and $\epsilon^{2}$. To the leading order in $\epsilon,$
i.e., $O(\epsilon^{0}),$ the KM system is 
\begin{gather}
\frac{\partial F_{s}^{(0)}}{\partial t}+\boldsymbol{v}\cdot\frac{\partial F_{s}^{(0)}}{\partial\boldsymbol{x}}+\left(\frac{q}{m}\right)_{s}\left(-\nabla\phi+\frac{1}{c}\boldsymbol{v}\times\boldsymbol{B}_{0}\right)\cdot\frac{\partial F_{s}^{(0)}}{\partial\boldsymbol{v}}=0,\label{eq:V}\\
\nabla^{2}\phi=-4\pi\sum_{s}q_{s}\int F_{s}^{(0)}d\boldsymbol{v},\label{eq:P-1}\\
\boldsymbol{E}_{l}^{(0)}\equiv-\nabla\phi,
\end{gather}
which is indeed the KP system. Higher-order equations can be derived
in a straightforward manner. 

For present purposes, we only need the first-order equation for the
first-order magnetic field $\boldsymbol{B}^{(1)}$,
\begin{equation}
\nabla\times\boldsymbol{B}^{(1)}=\frac{4\pi}{c}\sum_{s}q_{s}\int vF_{s}^{(0)}d\boldsymbol{v}+\frac{1}{c}\frac{\partial\boldsymbol{E}{}_{l}^{(0)}}{\partial t}.\label{eq:B1}
\end{equation}
Note that $\boldsymbol{B}^{(1)}$ is determined by the leading-order
fields $F_{s}^{(0)}$ and $\boldsymbol{E}{}_{l}^{(0)}$ due to the
fact that $v/c\sim\omega/ck\sim\epsilon\ll1.$ Even though $\boldsymbol{B}^{(1)}$
does not enter the governing equations for the KP system, i.e., Eqs.\,\eqref{eq:V}
and \eqref{eq:P-1}, it will enter the leading order energy conservation
law for the KP system. Starting from the exact energy conservation
law for the KM system, i.e., Eq.\,(8), we retain all the leading-order
terms to obtain
\begin{equation}
\frac{\partial}{\partial t}\left[\frac{(\nabla\phi)^{2}}{8\pi}+\sum_{s}\int F_{s}^{(0)}\frac{m_{s}\boldsymbol{v}^{2}}{2}d^{3}\boldsymbol{v}\right]+\nabla\cdot\left[\frac{-c\nabla\phi\times\boldsymbol{B}^{(1)}}{4\pi}+\sum_{s}\int F_{s}^{(0)}\frac{m_{s}\boldsymbol{v}^{2}}{2}\boldsymbol{v}d^{3}\boldsymbol{v}\right]=0.\label{eq:EC}
\end{equation}
The term $-c\nabla\phi\times\boldsymbol{B}^{(1)}/4\pi$ is the Poynting
flux due to the leading-order longitudinal electric field $\boldsymbol{E}_{l}^{(0)}$
and the first-order magnetic field $\boldsymbol{B}^{(1)}$, and it
must be included in the leading-order energy equation, because $ck/\omega\sim1/\epsilon$
raises the order of this term by one. Since $\boldsymbol{B}^{(1)}$
is uniquely determined by $F_{s}^{(0)}$ and $\boldsymbol{E}{}_{l}^{(0)}$
through Eq.\,\eqref{eq:B1}, the Poynting flux term can be expressed
as
\begin{align}
\nabla\cdot\left[\frac{-c\nabla\phi\times\boldsymbol{B}^{(1)}}{4\pi}\right]=\frac{c}{4\pi}\nabla\phi\cdot\nabla\times\boldsymbol{B}^{(1)}\nonumber \\
=\left[\sum_{s}q_{s}\int vF_{s}^{(0)}d\boldsymbol{v}-\frac{1}{4\pi}\nabla\phi_{,t}\right]\cdot\nabla\phi=\nabla & \cdot\left(\sum_{s}q_{s}\phi\int F_{s}^{(0)}\boldsymbol{v}d^{3}\boldsymbol{v}-\frac{1}{4\pi}\phi\nabla\phi_{,t}\right),\label{eq:KP-EC0}
\end{align}
where the continuity equation derived from Eq.\,\eqref{eq:V} has
been used. Finally, the leading-order energy conservation equation
is 
\begin{align}
 & \frac{\partial}{\partial t}\left[\frac{(\nabla\phi)^{2}}{8\pi}+\sum_{s}\int F_{s}^{(0)}\frac{m_{s}\boldsymbol{v}^{2}}{2}d^{3}\boldsymbol{v}\right]+\nonumber \\
 & \frac{\partial}{\partial\boldsymbol{x}}\cdot\left(\sum_{s}\int F_{s}^{(0)}\frac{m_{s}\boldsymbol{v}^{2}}{2}\boldsymbol{v}d^{3}\boldsymbol{v}+\sum_{s}q_{s}\phi\int F_{s}^{(0)}\boldsymbol{v}d^{3}\boldsymbol{v}-\frac{1}{4\pi}\phi\nabla\phi_{,t}\right)=0,\label{eq: KP-EC}
\end{align}
which is identical to Eq.\,\eqref{KM-EC}, if $F_{s}^{(0)}$ is identified
with $F_{s}.$ This demonstrates that the energy conservation derived
from the field theoretical approach for the KP system is not only
more rigorous in mathematical treatment, but also more correct in
physics content than the simple approach of replacing $\boldsymbol{E}$
by $-\nabla\phi$ and $\boldsymbol{B}$ by $\boldsymbol{B}_{0}$ in
the energy conservation law for the KM system. A similar argument
applies to the momentum conservation law for the KP system, i.e.,
Eq.\,\eqref{KM-M}. 
\begin{acknowledgments}
This research was supported by the CAS Program for Interdisciplinary
Collaboration Team, the JSPS-NRF-NSFC A3 Foresight Program in the
field of Plasma Physics (NSFC-11261140328), and the U.S. Department
of Energy (DE-AC02-09CH11466). 
\end{acknowledgments}

%

\end{document}